# Full susceptibility tensor for localized spin models

# with *S*=1, 3/2, 2, 5/2 and with rhombic anisotropy

R. Pełka[1]

H. Niewodniczański Institute of Nuclear Physics, Polish Academy of Sciences, Radzikowskiego 152, 31-342 Kraków, Poland

## Abstract

A general discussion of the simulation procedure of the full susceptibility tensor and isothermal magnetization pseudovector for compounds comprising weakly-interacting magnetic centers is presented. A single-crystal-sample as well as a powder-sample case are considered. The procedure is used to obtain explicit expressions for the full susceptibility tensor for spins *S*=1, 3/2, 2, and 5/2 for non-vanishing rhombic local anisotropy and any form of spectroscopic tensor.

[1] e-mail: robert.pelka@ifj.edu.pl

## 1. Introduction

The susceptibility and isothermal magnetization represent two widely used characteristics of magnetic materials. They carry the information on how the elementary magnetic moments associated with building units of a given compound respond to the external magnetic field. This response is dependent on the interaction pattern between the moments themselves and between them and other degrees of freedom present in the compound. The magnetic measurements of these properties together with the theoretical analysis of the ensuing data are to provide insight into the corresponding intra- and intermolecular interactions. It is therefore crucial to develop new experimental techniques on the one hand and refine theoretical apparatus on the other. The need for such considerations is strongly suggested by the developments in the field of coordination compounds [1-3], in general, and in that of isolated polynuclear d- or f-electronic systems displaying considerable magnetic anisotropies [4-9], in particular.

Although comprehensive reviews of these issues are available [10,11], this contribution includes some useful complementary results. It concerns the discussion of the procedure referred to in [10] as the generalized van Vleck formula. It also provides exact formulae for the full susceptibility tensor for localized spin models with $S$=1, 3/2, 2, and 5/2 with both axial and rhombic zero-field splitting term and for any spectroscopic tensor $g_{ij}$. In [10] and [11] analytical expressions are given only for the case of nonvanishing axial anisotropy and isotropic spectroscopic tensor, and the general case taking into account the rhombic zero-field splitting term is treated only numerically. Exact results for this lowest symmetry case is also missing in the seminal book on molecular magnetism by Otto Kahn [12]. Furthermore, the reviews [10,11] do not discuss the issue of the calculation of the full susceptibility tensor, which is of major importance if one wants to correctly analyze the experimental single-crystal data. The approach presented in this paper enables one to obtain all terms of the susceptibility tensor.

## 2. Theoretical background

Let us consider a magnetic material which consists of magnetic centers distributed over the crystal lattice in such a way that each has the same spatial orientation. Moreover, let us assume that the interaction between the centers is negligibly small, thus the theoretical analysis may be confined to a single center. The microscopic properties of that center are determined by the Hamiltonian which is split into two components, i.e. $\hat{H} = \hat{H}_0 + \hat{V}$. The first component accounts for the interaction of the center with its nearest environment and with an external magnetic field $\boldsymbol{H_0}$ corresponding to the field fixed during the measurement. The other component is the Zeeman coupling of the center to the magnetic field $\boldsymbol{H}$ which plays the role of the sampling field. While the latter term is thought to be the perturbation, the former one represents the unperturbed system. The standard quantum-mechanical scheme of the perturbative calculation [13] yields consecutive corrections to the energies of $E_n^0(\boldsymbol{H_0}) = \langle n, i_n | \hat{H}_0 | n, i_n \rangle$ of eigenstates $|n, i_n\rangle$ of $\hat{H}_0$ ( $n = 1, \ldots, d$ ; $i_n = 1, \ldots, f_n$, where $f_n$ is the degeneracy degree of the $n$-th state). The first and second order corrections may be written as diagonal elements of the following operators

$$\hat{C}^1 = \hat{V} \tag{2.1}$$

$$\hat{C}_n^2 = \sum_{m \neq n} \sum_{i_m = 1}^{f_m} \frac{\hat{V} | m, i_m \rangle \langle m, i_m | \hat{V}}{E_n^0 - E_m^0} \tag{2.2}$$

respectively, where states $|n, i_n\rangle$ are assumed to be chosen in such a way that operator $\hat{C}^1$ is diagonal in each degenerate subspace, and if it vanishes in any subspace, the corresponding second order operator $\hat{C}_n^2$ has there the diagonal form. So, in a generic case, calculating the energy corrections involves diagonalizing procedures in every order of the perturbative scheme. The principles of statistical physics imply the following formulae for the magnetization and susceptibility components

$$M_i = \frac{1}{Z}\sum_{n,i_n}\left(-\frac{\partial E_{n,i_n}}{\partial H_i}\right)\exp(-\beta E_{n,i_n}) \tag{2.3}$$

$$\beta^{-1}\chi_{ij} = \sum_{n,i_n}\left(\frac{\partial E_{n,i_n}}{\partial H_i}\frac{\partial E_{n,i_n}}{\partial H_i} - \beta^{-1}\frac{\partial^2 E_{n,i_n}}{\partial H_i \partial H_j}\right)\frac{\exp(-\beta E_{n,i_n})}{Z} - M_i M_j \tag{2.4}$$

$$Z = \sum_{n,i_n}\exp(-\beta E_{n,i_n}) \tag{2.5}$$

where all the functions $E_{n,i_n}(\boldsymbol{H_0}, \boldsymbol{H})$ and their derivatives are evaluated at $\boldsymbol{H} = (0,0,0)$, indices $i, j \in \{x, y, z\}$, and $\beta = 1/kT$. Remembering that perturbation operator $\hat{V}$ is linear in the sampling field $\boldsymbol{H}$, i.e. $\hat{V} = \hat{V}_i H_i$, where the Einstein summation convention is used, the straightforward calculation yields

$$M_i = \frac{1}{Z}\sum_n \exp(-\beta E_n^0)\sigma_{i,n}^1 \tag{2.6}$$

$$\beta^{-1}\chi_{ij} = \frac{1}{Z}\sum_n \exp(-\beta E_n^0)(\sigma_{ij,n}^2 - \beta^{-1}\sigma_{ij,n}^3) - M_i M_j \tag{2.7}$$

where $Z = \sum_n f_n \exp(-\beta E_n^0)$ and the sigmas have the following forms

$$\sigma_{i,n}^1 = \sum_{i_n}\langle n, i_n|\hat{V}_i|n, i_n\rangle \tag{2.8}$$

$$\sigma_{ij,n}^2 = \sum_{i_n}\langle n, i_n|\hat{V}_i|n, i_n\rangle\langle n, i_n|\hat{V}_i|n, i_n\rangle \tag{2.9}$$

$$\sigma_{ij,n}^3 = 2\,\mathrm{Re}\sum_{i_n}\sum_{n\neq m, i_m}\frac{\langle n, i_n|\hat{V}_i|m, i_m\rangle\langle m, i_m|\hat{V}_j|n, i_n\rangle}{E_n^0 - E_m^0} \tag{2.10}$$

Let us stress that the procedure defined by Eqs. (2.6-10) is exact and does not entail any approximations contrary to what the quantum mechanical expansion for the energy of states might suggest. Sums $\sigma^1_{i,n}$ and $\sigma^3_{ij,n}$ in Eqs. (2.8) and (2.10), respectively, can be interpreted as the trace in the $n$-th subspace of degenerate states, whereas sum $\sigma^2_{ij,n}$ cannot. As the trace is independent of the choice of the basis in the state space, the calculation of the magnetization components does not require the block diagonalization of $\hat{C}^1$. This operation is yet necessary in the calculation of the susceptibility components.

It is instructive to calculate the magnetic properties for the simplest possible system of a noninteracting spin $S$ in a nonzero external magnetic field $\boldsymbol{H_0}$. The pertinent Hamiltonian reads

$$\hat{H} = \mu_0 \hat{S} \cdot (\boldsymbol{H_0} + \boldsymbol{H}) \tag{2.11}$$

with $\mu_0 = g\mu_B$, where $\hat{S}$ is the spin operator vector, $g$ - the spectroscopic Landé factor, and $\mu_B$ - the Bohr magneton. In this simple case the total Hamiltonian can be diagonalized explicitly to yield the eigenvalues enumerated by the magnetic quantum number $M = -S, -S+1, \ldots, S-1, S$: $E_M(\boldsymbol{H_0}, \boldsymbol{H}) = \mu_0 M |\boldsymbol{H_0} + \boldsymbol{H}|$, of which the relevant derivatives can be obtained directly without resorting to the perturbative scheme. The magnetization and molar susceptibility tensor of that system found using Eqs. (2.3-5) read

$$\boldsymbol{M} = \mu_0 B_S(\beta\mu_0 S|\boldsymbol{H_0}|)\frac{\boldsymbol{H_0}}{|\boldsymbol{H_0}|} \tag{2.12}$$

$$\beta^{-1}\chi_{ij} = N_A \mu_0^2 S^2 \left[ w_1(\beta, |\boldsymbol{H_0}|)\delta_{ij} + w_2(\beta, |\boldsymbol{H_0}|)\frac{H_{0i}H_{0j}}{|\boldsymbol{H_0}|^2} \right] \tag{2.13}$$

where

$$w_1(\beta, |\boldsymbol{H_0}|) = \frac{B_S(\beta\mu_0 S|\boldsymbol{H_0}|)}{\beta\mu_0 S|\boldsymbol{H_0}|} \tag{2.14}$$

$$w_2(\beta,|\boldsymbol{H_0}|) = \frac{S+1}{S} - \frac{1}{S}\coth\left(\frac{1}{2}\beta\mu_0|\boldsymbol{H_0}|\right)B_S(\beta\mu_0 S|\boldsymbol{H_0}|) - \frac{B_S(\beta\mu_0 S|\boldsymbol{H_0}|)}{\beta\mu_0 S|\boldsymbol{H_0}|} - B_S^2(\beta\mu_0 S|\boldsymbol{H_0}|) \tag{2.15}$$

where $N_A$ is the Avogadro number, and $B_S(x)$ is the Brillouin function. The susceptibility tensor may be shown to be diagonal in the frame $(\boldsymbol{e}_{\|}, \boldsymbol{e}_{\perp 1}, \boldsymbol{e}_{\perp 2})$ where $\boldsymbol{e}_{\|} = \boldsymbol{H_0}/|\boldsymbol{H_0}|$, and $\boldsymbol{e}_{\perp i} \cdot \boldsymbol{e}_{\|} = 0$ ($i = 1,2$). The corresponding eigenvalues are given by the following formulae

$$\chi_{\|} = N_A\beta[w_1(\beta,|\boldsymbol{H_0}|) + w_2(\beta,|\boldsymbol{H_0}|)] \tag{2.16}$$

$$\chi_{\perp} = N_A\beta w_1(\beta,|\boldsymbol{H_0}|) \tag{2.17}$$

The second eigenvalue is twofold degenerate. Figure 1 shows temperature dependence of the susceptibility eigenvalues given by Eqs. (2.16) and (2.17) for five spin values $S$=1/2, 1, 3/2, 2, 5/2, and $g = 2.0$ in the external magnetic field of the magnitude $H_0 = 10^3$ Oe typical for genuine experimental conditions. For comparison it also shows the triply degenerate eigenvalue of the susceptibility tensor in the vanishing external magnetic field ($|\boldsymbol{H_0}| \to 0$). It can be seen that the spectrum of the susceptibility tensor changes considerably if a nonzero external magnetic field is present. Eigenvalue $\chi_{\|}$ vanishes whereas eigenvalue $\chi_{\perp}$ is finite in the zero-temperature limit. In the zero-field limit the single triply degenerate eigenvalue $\chi_0 = N_A\mu_B^2 g^2\beta S(S+1)/3$ diverges for $T \to 0$.

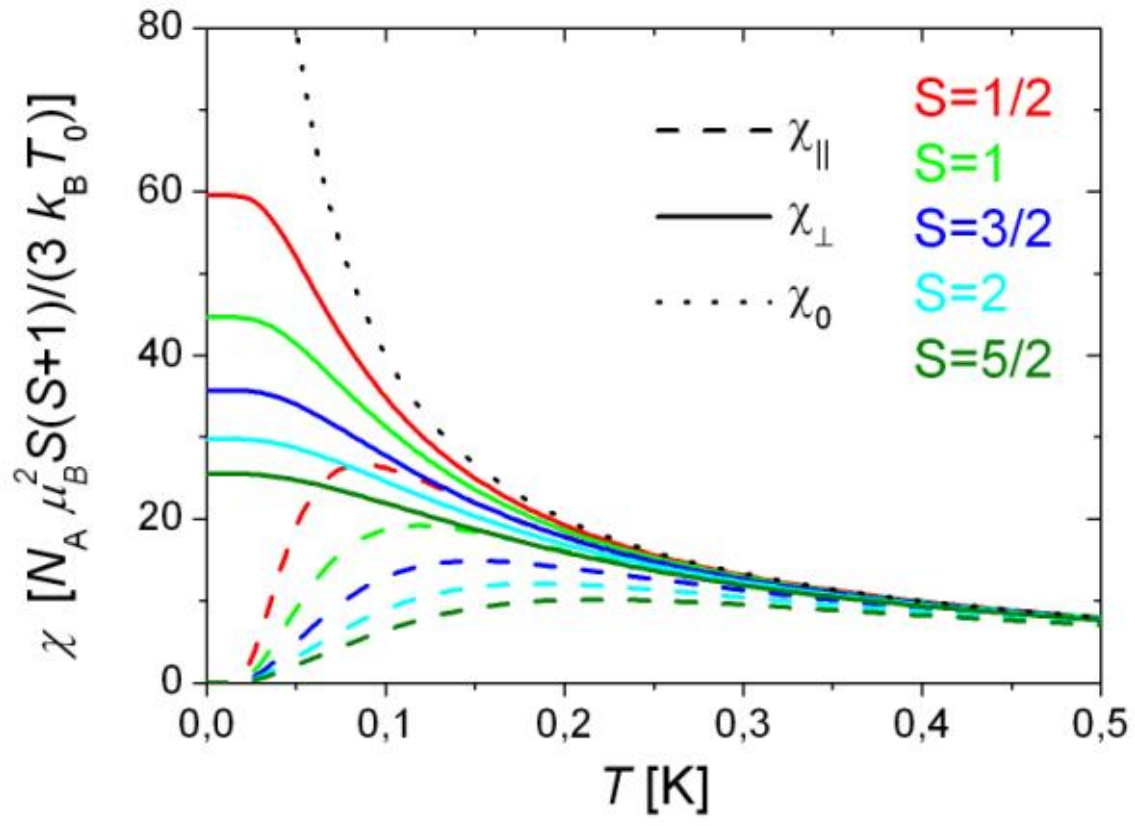

**Fig. 1:** Temperature dependence of eigenvalues $\chi_{\|}$(dashed line) and $\chi_{\perp}$(solid line) of the susceptibility tensor for an isolated spin system with $S$=1/2, 1, 3/2, 2, 5/2, and $g = 2.0$ in the external magnetic field of $10^3$ Oe. The dotted line shows the triply degenerate eigenvalue of the susceptibility tensor $\chi_0$ in the zero-field limit. Note that the susceptibility units are $S$-dependent and have been chosen so for all the curves to coincide in high temperature limit.

To conclude this section let us generalize the previous result to encompass the case where there are several weakly-interacting centers in the crystallographic cell differing in orientation from each other. Firstly, one needs to determine for each center its local coordinate frame $F_k$ ($k = 1,\ldots,N$, $N$ – the number of centers in the crystallographic cell) in the way consistent with the corresponding coordination sphere and the existing conventions. Let $F$ denote the orthonormal frame related to the crystallographic axes and $O_k$ - the orthogonal transformations from $F$ to $F_k$ (the transformations are defined by the matrix equation $E_k = EO_k$, where $E$ denotes the 3×3 matrix whose columns are the basis vectors of frame $F$ and $E_k$ is the 3×3 matrix whose columns are the basis vectors of local coordinate frame $F_k$). The external magnetic field $\boldsymbol{H_0}$ (coordinates in frame $F$) must be transformed to the local frames: $\boldsymbol{H}_k = \det O_k O_k^{\mathrm{T}} \boldsymbol{H_0}$. Then the calculations are performed for each center separately in its local frame to obtain $\boldsymbol{M}_k$ and $\chi_k$. The total magnetization and susceptibility will then be given by the following sums

$$\boldsymbol{M}_{\text{total}} = \sum_{k=1}^{N} \det O_k O_k \boldsymbol{M}_k \tag{2.18}$$

$$\boldsymbol{\chi}_{\text{total}} = \sum_{k=1}^{N} O_k \boldsymbol{\chi}_k O_k^{\mathrm{T}} \tag{2.19}$$

In the case where the external magnetic field $\boldsymbol{H_0}$ is practically zero and all the centers are isomorphic, the calculation of the total susceptibility requires only the calculation in one local frame as $\boldsymbol{\chi}_k(\boldsymbol{H_0} = \boldsymbol{0}) = \boldsymbol{\chi}_1(\boldsymbol{H_0} = \boldsymbol{0})$ .

## 3. Experimental background

In the previous section the procedure allowing the calculation of the magnetization and the susceptibility of the crystalline magnetic material was presented. Yet, what is calculated must often be compared to the experimental output. The current section is devoted to the issue of what quantities are available experimentally and how they correspond to the theoretical ones.

The reasonable assumption concerning magnetic measurements will be that the quantity measured corresponds to the magnetization of the sample induced by the external magnetic field $\boldsymbol{H_0}$ . Its mathematical counterpart may be represented by the formula

$$m = \frac{\boldsymbol{M} \cdot \boldsymbol{H_0}}{|\boldsymbol{H_0}|} \tag{3.1}$$

For finite values of the magnetic field and fixed temperature one obtains the isothermal magnetization, corresponding for a single-crystal sample of the multicenter compound to the following sum (cf. Eq. (2.18) and the text above)

$$m = \sum_{k=1}^{N} \frac{\boldsymbol{M}_k \cdot \boldsymbol{H}_k}{|\boldsymbol{H_0}|} \tag{3.2}$$

Independent orientations of the external field $\boldsymbol{H_0}$ will yield the corresponding Carthesian components of the magnetization pseudovector. If one takes the background field to be small $\boldsymbol{H_0} \to \delta\boldsymbol{H_0}$, one can expand the magnetization in Eq. (3.1) around zero to obtain

$$M_i = \chi_{ij}(\boldsymbol{0})\delta H_{0j} + O(\delta H_0^2) \tag{3.3}$$

and consequently

$$\frac{m}{\delta H_0} \approx \frac{\delta H_{0i}}{\delta H_0} \chi_{ij} \frac{\delta H_{0j}}{\delta H_0} \tag{3.4}$$

where $\delta H_0 = |\delta \boldsymbol{H_0}|$ and the Einstein summation convention applies. Thus one obtains the quantity proportional to the diagonal components of the susceptibility tensor. It is easy to see that this particular experimental setup defined by Eq. (3.1) does not allow for the determination of the off-diagonal terms.

For powder samples the averaging over all possible orientations of the magnetic centers or, equivalently, over all possible directions of the external magnetic field must be carried out. Such an averaging can be performed analytically only for the most simple cases. In a generic case one has to resort to numerical computation. One of possible selections of orientations of the external magnetic field $\boldsymbol{H_0}$ is that corresponding to a grid of points homogenously filling the rectangle $[0,1]\times[0,2\pi)$ in the coordinate plane $(\cos\theta, \varphi)$, where the couple of spherical angles $(\theta, \varphi)$ provide the most natural parametrization of the field direction. This particular choice was invented to represent the opposite directions of the field only ones and involves angle steps defined by $\Delta(\cos\theta) = 1/K$ and $\Delta\varphi = \pi/(2K)$, where $K = 1,2,\ldots$. The ensuing set of points is a sum of the three following disjoint angle sets: (1) $\{(0,0)\}$, (2) $\left\{\left(\frac{\pi}{2}, \frac{\pi}{2K}(i-1)\right); i = 1,\ldots,2K\right\}$, and (3) $\left\{\left(\arccos\left(\frac{i}{K}\right), \frac{\pi}{2K}(j-1)\right); i=1,\ldots,K-1, j=1,\ldots,4K\right\}$. It is easy to see that the total number of points (orientations) is then $P=4K^2-2K+1$. Thus for a one-center compound one has to carry out the following averaging

$$m \approx \frac{1}{P}\sum_{l=1}^{P}\frac{\boldsymbol{M}(\boldsymbol{H}_l)\cdot\boldsymbol{H}_l}{|\boldsymbol{H}_l|} \tag{3.5}$$

where $|\boldsymbol{H}_l|=|\boldsymbol{H}_0|$ and field vectors $\boldsymbol{H}_l$ represent the different equally-weighted field orientations. In the simplest case, i.e. for $K$=1, Eq. (3.5) yields $m \approx (M_x+M_y+M_z)/3$. However, this approximation will be demonstrated later to fail to provide a reliable estimate. The averaging for the zero-field susceptibility, see Eq. (3.4), yields the widely-used formula [10-12]

$$\frac{m}{\delta H_0}=\frac{1}{4\pi}\int \mathrm{d}\Omega \frac{\delta\boldsymbol{H}_0^{\mathrm{T}}(\Omega)\boldsymbol{\chi}(\boldsymbol{0})\delta\boldsymbol{H}_0(\Omega)}{\delta H_0^2}=\frac{1}{3}\mathrm{Tr}\boldsymbol{\chi}(\boldsymbol{0}) \tag{3.6}$$

## 4. Full susceptibility tensor for small spin values

The so called generalized van Vleck formalism summarized in Section 2 can be used to obtain typical characteristics of magnetic systems, i.e. magnetic susceptibility tensor and isothermal magnetization pseudovector. These can be calculated numerically for an arbitrary magnetic system taking into account an arbitrary value of the external magnetic field. However, there is an important class of systems where the calculations of magnetic susceptibility tensor in the zero external magnetic field can be carried out analytically. To that class belong systems of localized spins immersed in an anisotropic environment. The pertinent phenomenological Hamiltonian for those systems has the following form

$$\hat{H}=D\hat{S}_z^2+E(\hat{S}_x^2-\hat{S}_y^2)+\mu_B\hat{\boldsymbol{S}}\cdot\boldsymbol{g}\cdot\boldsymbol{H} \tag{4.1}$$

The first two terms in Eq. (4.1) account for the axial and rhombic zero-field splitting, respectively, with axial (*D*) and rhombic (*E*) zero-field splitting parameters [12]. The last term corresponds to the Zeeman coupling of the spin with the external magnetic field. Let us note that we do not assume the diagonal form of the spectroscopic tensor $\boldsymbol{g}$. In the following subsections we include analytical formulae for the susceptibility tensor for an array of the lowest spin values found with Eqs. (2.6-10).

### 4.1 The case of *S*=1

The spectrum of the zeroth order Hamiltonian $\hat{H}_0$ is nondegenerate. The corresponding eigenvalues and eigenvectors are listed in Section 6.1. The susceptibility tensor, obtained using Eqs. (2.6-10) and (6.1.1-2) reads

$$\beta^{-1}\chi_{ij} = \frac{2\mu_B^2}{e^{\beta D} + 2\cosh(\beta E)} \left\{ \exp\left[\frac{1}{2}\beta(D-E)\right] \frac{\sinh\left[\frac{1}{2}\beta(D+E)\right]}{\frac{1}{2}\beta(D+E)} g_{xi}g_{xj} + \exp\left[\frac{1}{2}\beta(D+E)\right] \frac{\sinh\left[\frac{1}{2}\beta(D-E)\right]}{\frac{1}{2}\beta(D-E)} g_{yi}g_{yj} + \frac{\sinh(\beta E)}{\beta E} g_{zi}g_{zj} \right\} \quad (4.1.1)$$

Figure 2 shows the susceptibility components calculated numerically in the external field $\boldsymbol{H_0} = [1,1,1]\,\mathrm{kOe}$ for *D*=-10 K, *E*=5 K, and $\boldsymbol{g} = \mathrm{diag}(2,2,2)$. The numerical calculations are meant to verify the correctness of the analytical results. The numerical procedure was prepared in the *Mathematica 7.0* environment in the simplified yet sufficient version where the energy spectrum is non-degenerate. In this and all the remaining cases the value of the external magnetic field $\boldsymbol{H_0}$, required by the numerical procedure, was chosen to be small so that the susceptibility values obtained with the derived formulae and those calculated numerically did not visibly differ except at very low temperatures. The numerical procedure yields the full susceptibility tensor and the mean susceptibility ($\chi_{mean}$) curves shown in plots correspond to one-third of the trace of that tensor. From Fig. 2 it can be seen that the numerical calculation is consistent with the analytical result given by Eq. (4.1.1) shown by solid lines. The inset of Fig 2 shows the temperature dependence of the off-diagonal components of the susceptibility tensor that appear due to the presence of the external magnetic field. The magnetization was simulated in the three independent spatial directions. Moreover, the magnetization for the powder sample was calculated for $K$=1, 2, 3, 5, and 10. The results are depicted in Fig. 3. Let us note that the simplest approximation with $K$=1 deviates considerably from that obtained with $K$=10 involving $P$=381 different field orientations.

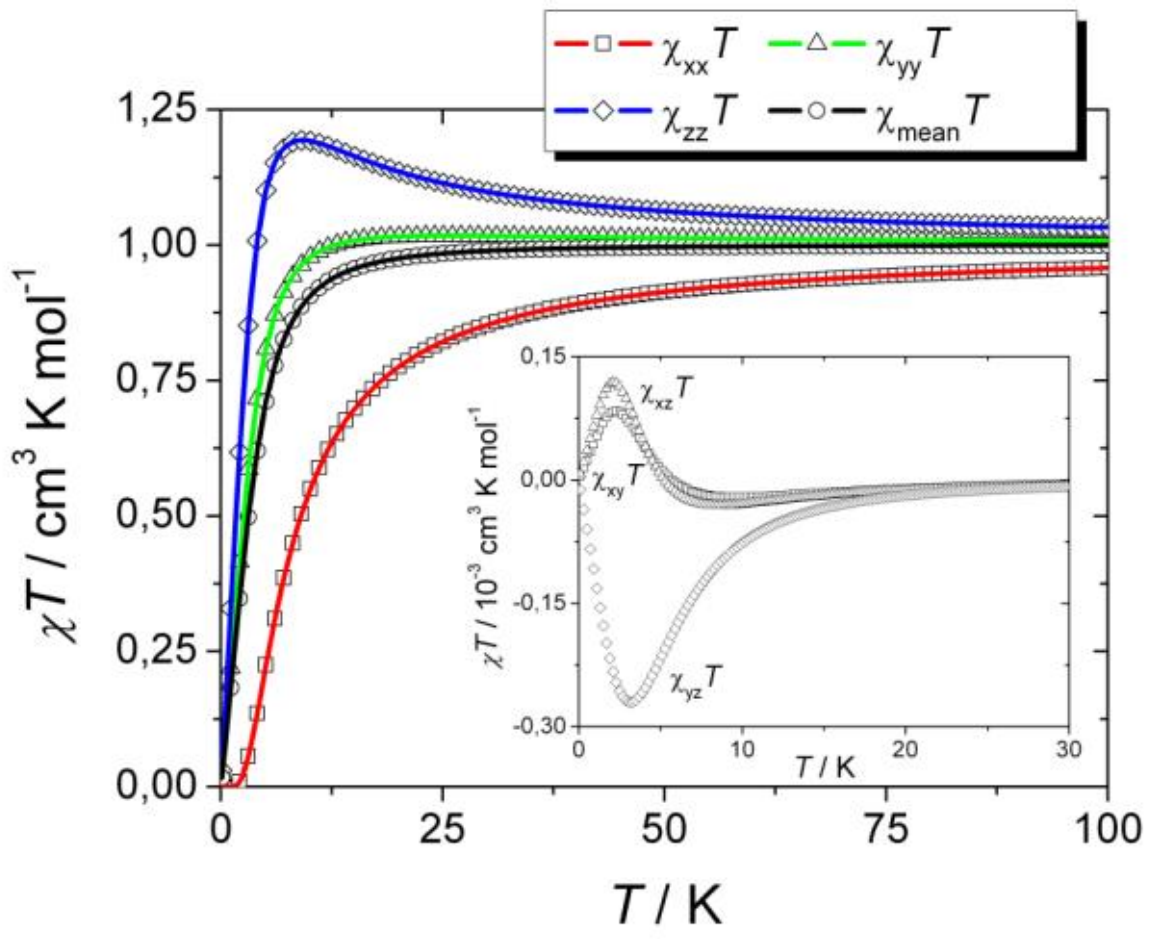

**Fig. 2:** Components of the susceptibility tensor simulated in the model defined by Hamiltonian given in Eq. (4.1) for $S$=1, $D$=-10 K, $E$=5 K and $\boldsymbol{g} = \mathrm{diag}(2,2,2)$. The solid lines show the diagonal components of the zero-field susceptibility calculated according to Eq. (4.1.1). The black solid line shows the mean (powder sample) susceptibility calculated as one third of the trace of the tensor given by Eq. (4.1.1). The Inset shows the off-diagonal components due to the presence of the external magnetic field $\boldsymbol{H_0} = [1,1,1]$ kOe.

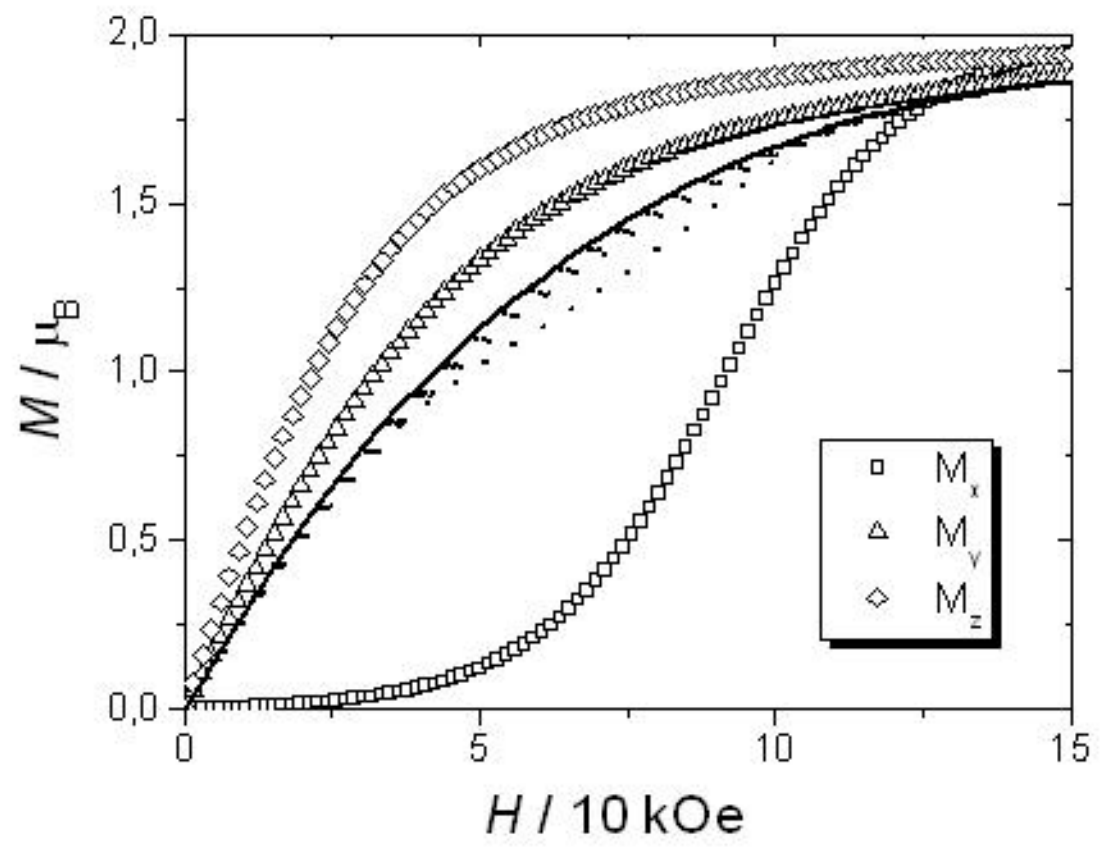

**Fig. 3:** Components of the magnetization simulated in the model defined by Hamiltonian given in Eq. (4.1) for $S$=1, $T$=2 K, $D$=-10 K, $E$=5 K and $\boldsymbol{g} = \mathrm{diag}(2,2,2)$. The dotted curves denote the powder sample magnetizations obtained with $K$=1, 2, 3, 5 corresponding to $P$=3, 13, 31, 91 field orientations, respectively. The solid line shows the powder sample magnetization for $K$=10 ($P$=381). A considerable difference can be observed between the curves with $K$=1 and $K$=10.

### 4.2 The case of *S*=3/2

For the odd multiples of spin ½ the well-known Kramers theorem applies. The crystal field, which originates from electrostatic interactions of valence electrons with the crystal environment (coordination sphere), cannot fully remove the degeneracy of the spectrum for such systems. The spectrum consists of a number of Kramers doublets, which can be split only by applying an external magnetic field. The spectrum of the zeroth order Hamiltonian $\hat{H}_0$ in the present case comprises a pair of Kramers doublets. The corresponding eigenvalues and eigenvectors are listed in Section 6.2. One can see that the eigenvectors depend on the sampling field. This feature was absent for the case of the integer spin $S$=1 and is characteristic for spins equal to odd multiples of ½. Even in the limit of $\boldsymbol{H} \to \boldsymbol{0}$ it introduces a nontrivial dependence of the susceptibility tensor on the direction of the sampling field. Using Eqs. (2.6-10) and the appropriately chosen eigenvectors given by Eqs. (6.2.2-5) one arrives at the final formula for the zero-field susceptibility tensor

$$\beta^{-1}\chi_{ij}(\boldsymbol{h}) = \frac{1}{4}\mu_B^2\left\{\frac{1}{2}\left[\frac{A_{2ik}A_{2jl}h_k h_l}{A_{2kl}h_k h_l} + \frac{A_{1ik}A_{1jl}h_k h_l}{A_{1kl}h_k h_l}\right]\right.$$
$$+\frac{1}{2}\left[\frac{A_{2ik}A_{2jl}h_k h_l}{A_{2kl}h_k h_l} - \frac{A_{1ik}A_{1jl}h_k h_l}{A_{1kl}h_k h_l}\right]\tanh(\beta\Delta) \tag{4.2.1}$$
$$\left.+\frac{3}{\Delta^2}[(D+E)^2 g_{xi}g_{xj} + (D-E)^2 g_{yi}g_{yj} + 4E^2 g_{zi}g_{zj}]\frac{\tanh(\beta\Delta)}{\beta\Delta}\right\}$$

where quantities $A_1$ and $A_2$ are given by the following formula

$$A_{1/2ij} = \left(1 \mp \frac{D-3E}{\Delta}\right)^2 g_{xi}g_{xj} + \left(1 \mp \frac{D+3E}{\Delta}\right)^2 g_{yi}g_{yj} + \left(1 \pm \frac{2D}{\Delta}\right)^2 g_{zi}g_{zj} \tag{4.2.2}$$

and versor $\boldsymbol{h} = \boldsymbol{H}/|\boldsymbol{H}|$ encodes the orientation of the sampling field. The presence of $\boldsymbol{h}$ in the final formula for the susceptibility tensor calls for caution while calculating powder and single-crystal magnetic susceptibility. To duly account for the powder sample signal it is no more sufficient to calculate the susceptibility tensor once for a fixed sampling field orientation $\boldsymbol{h}$ and then take one third of its trace as this will give an incorrect result. To obtain a correct result one has to repeat the calculations for an array of sampling field orientations, like those assumed for the calculation of powder magnetization, and next take an average. In fact that averaging procedure can be performed exactly without resorting to its discretization. The corresponding result will be presented and discussed later in this section. The same applies, when a single crystal with several isomorphic spin centers in a unit cell is considered. We cannot transform the susceptibility tensor calculated for one center to obtain that tensor for another center differing from the first one in the spatial orientation, because the relative orientation of the external magnetic field changes from center to center. The calculations should be performed for each center separately and only then the results should be summed to yield the total susceptibility of a unit cell.

In many practical cases only the three linearly independent configurations of the sampling field $\boldsymbol{h}_x = [1,0,0]$, $\boldsymbol{h}_y = [0,1,0]$, and $\boldsymbol{h}_z = [0,0,1]$ and the corresponding diagonal terms of the susceptibility tensor $\chi_{xx}(\boldsymbol{h}_x)$, $\chi_{yy}(\boldsymbol{h}_y)$, and $\chi_{zz}(\boldsymbol{h}_z)$ are considered. On those assumptions Eq. (4.2.1) simplifies, yielding the following compact formula

$$\begin{aligned}\frac{\beta^{-1}\chi_{ii}(\boldsymbol{h}_i)}{\mu_B^2} &= g_{xi}^2\left[\frac{5}{4}-\frac{3(D+E)^2}{4\Delta^2}\left(1-\frac{\tanh(\beta\Delta)}{\beta\Delta}\right)+\frac{D-3E}{2\Delta}\tanh(\beta\Delta)\right] \\ &+g_{yi}^2\left[\frac{5}{4}-\frac{3(D-E)^2}{4\Delta^2}\left(1-\frac{\tanh(\beta\Delta)}{\beta\Delta}\right)+\frac{D+3E}{2\Delta}\tanh(\beta\Delta)\right] \\ &+g_{zi}^2\left[\frac{5}{4}-\frac{3E^2}{\Delta^2}\left(1-\frac{\tanh(\beta\Delta)}{\beta\Delta}\right)-\frac{D}{\Delta}\tanh(\beta\Delta)\right]\end{aligned} \tag{4.2.3}$$

where index $i$ should be replaced with $x$, $y$, and $z$, respectively. Now, let us turn to carrying out the powder averaging of the susceptibility tensor. To this end, as pointed out above, we need to calculate an integral analogous to that given in Eq. (3.6), i.e.

$$\bar{\chi} = \frac{1}{4\pi}\int \mathrm{d}\Omega h_i(\Omega)\chi_{ij}(\Omega)h_j(\Omega) \tag{4.2.4}$$

where $\chi_{ij}(\Omega)$ is given by Eq. (4.2.1), $\boldsymbol{h}(\Omega)=[\sin\theta\cos\varphi,\sin\theta\sin\varphi,\cos\theta]$, and $(\theta,\varphi)$ is the couple of spherical angles parametrizing the sampling field direction. It is easy to show that for any $\Omega$-independent tensor $a_{ij}$ the following identity holds

$$\frac{1}{4\pi}\int \mathrm{d}\Omega h_i(\Omega)a_{ij}h_j(\Omega) = \frac{1}{3}(a_{xx}+a_{yy}+a_{zz}) \tag{4.2.5}$$

Using (4.2.4) and (4.2.5) one readily arrives at the formula for the susceptibility of the powder sample

$$\bar{\chi} = \frac{1}{3}[\chi_{xx}(\boldsymbol{h}_x)+\chi_{yy}(\boldsymbol{h}_y)+\chi_{zz}(\boldsymbol{h}_z)] \tag{4.2.6}$$

where $\chi_{ii}(\boldsymbol{h}_i)$ are given by Eq. (4.2.3). Let us stress that although the formula looks like one third of the trace of a susceptibility tensor, it is not the case as $\chi_{xx}(\boldsymbol{h}_x)$, $\chi_{yy}(\boldsymbol{h}_y)$, and $\chi_{zz}(\boldsymbol{h}_z)$ are the diagonal components of three different tensors corresponding to three different sampling field orientations. The same subtle situation will be present in Section 4.4 where the other half-integer spin $S$=5/2 will be discussed.

Figure 4 shows the susceptibility components calculated numerically in the external field $\boldsymbol{H_0} = [10,10,10]$ Oe for $D$=-10 K, $E$=5 K, and $\boldsymbol{g} = \mathrm{diag}(2,2,2)$. The numerical results are consistent with those obtained using the analytical formulae given by Eqs. (4.2.1) and (4.2.6) (shown by solid lines).

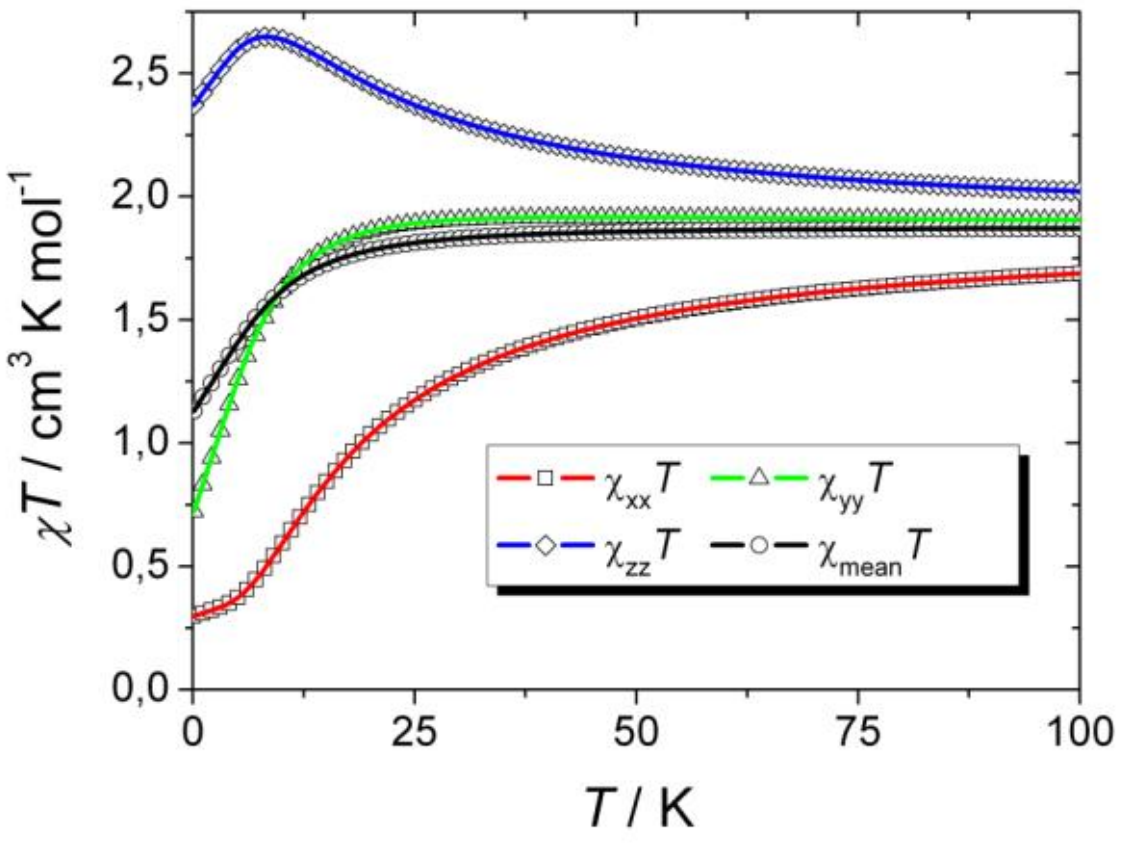

**Fig. 4:** Components of the susceptibility tensor simulated in the model defined by Hamiltonian given in Eq. (4.1) for $S$=3/2, $D$=-10 K, $E$=5 K and $\boldsymbol{g} = \mathrm{diag}(2,2,2)$. The solid lines show the diagonal components of the zero-field susceptibility calculated according to Eq. (4.2.1). The black solid line represents the mean (powder sample) susceptibility calculated according to Eq. (4.2.6). The numerical calculation was performed at the presence of the external magnetic field $\boldsymbol{H_0} = [10,10,10]$ Oe.

### 4.3 The case of $S$=2

The spectrum of the zeroth order Hamiltonian $\hat{H}_0$ is non-degenerate. The corresponding eigenvalues and eigenvectors are listed in Section 6.3. Using Eqs. (2.6-10) and (6.3.1-4) one arrives at the formula for the susceptibility tensor

$$\beta^{-1}\chi_{ij} = \frac{2\mu_B^2}{Z_1}\{C_1(\beta, D, E)g_{xi}g_{xj} + C_2(\beta, D, E)g_{yi}g_{yj} + C_3(\beta, D, E)g_{zi}g_{zj}\} \qquad (4.3.1)$$

where $Z_1 = \exp(-2\beta D) + 2\exp(\beta D)\cosh(3\beta E) + 2\cosh(2\beta\Delta)$, and functions $C_1$, $C_2$, and $C_3$ have the following form

$$\begin{aligned} C_1(\beta, D, E) &= \exp\left[-\frac{1}{2}\beta(D-3E)\right]\frac{\sinh\left[\frac{3}{2}\beta(D+E)\right]}{\frac{3}{2}\beta(D+E)} \\ &+ \left(2 - \frac{D-3E}{\Delta}\right)\exp\left[-\frac{1}{2}\beta(2\Delta - D + 3E)\right]\frac{\sinh\left[\frac{1}{2}\beta(2\Delta + D - 3E)\right]}{\frac{1}{2}\beta(2\Delta + D - 3E)} \\ &+ \left(2 + \frac{D-3E}{\Delta}\right)\exp\left[\frac{1}{2}\beta(2\Delta + D - 3E)\right]\frac{\sinh\left[\frac{1}{2}\beta(2\Delta - D + 3E)\right]}{\frac{1}{2}\beta(2\Delta - D + 3E)} \end{aligned} \qquad (4.3.2)$$

$$C_2(\beta, D, E) = C_1(\beta, D, -E) \qquad (4.3.3)$$

$$\begin{aligned} C_3(\beta, D, E) &= 2\left(1 + \frac{D}{\Delta}\right)\exp[-\beta(\Delta + D)]\frac{\sinh[\beta(\Delta - D)]}{\beta(\Delta - D)} \\ &+ 2\left(1 - \frac{D}{\Delta}\right)\exp[\beta(\Delta - D)]\frac{\sinh[\beta(\Delta + D)]}{\beta(\Delta + D)} \\ &+ \exp(\beta D)\frac{\sinh[3\beta E]}{3\beta E} \end{aligned} \qquad (4.3.4)$$

Figure 5 shows the susceptibility components calculated numerically in the external field $\boldsymbol{H_0} = [1,1,1]\,\mathrm{kOe}$ for $D$=-10 K, $E$=5 K, and $\boldsymbol{g} = \mathrm{diag}(2,2,2)$. The numerical results are consistent with the analytical result given by Eq. (4.3.1) (shown by solid lines).

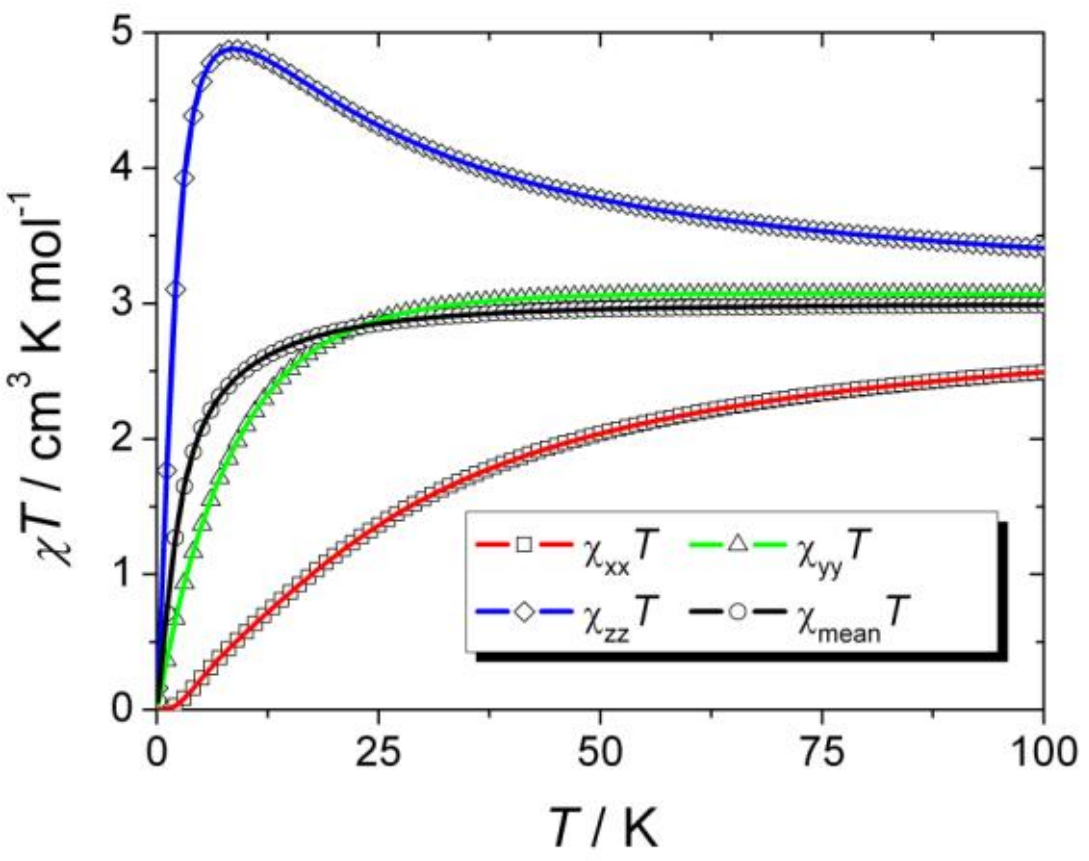

**Fig. 5:** Components of the susceptibility tensor simulated in the model defined by Hamiltonian given in Eq. (4.1) for $S$=2, $D$=-10 K, $E$=5 K and $\boldsymbol{g} = \mathrm{diag}(2,2,2)$. The solid lines show the diagonal components of the zero-field susceptibility calculated according to Eq. (4.3.1). The black solid line represents the mean (powder sample) susceptibility calculated as one third of the trace of the tensor given in Eq. (4.3.1). The numerical calculation was performed at the presence of the external magnetic field $\boldsymbol{H_0} = [1,1,1]\,\mathrm{kOe}$.

### 4.4 The case of *S*=5/2

The spectrum of the zeroth order Hamiltonian $\hat{H}_0$ consists of three Kramers doublets. Let us denote the corresponding eigenvalues $E_n^{(0)}$ by $\varepsilon_n$ ($n$=1,2,3), respectively, where

$$\varepsilon_1 = \frac{4\sqrt{7}}{3}\Delta\cos\left(\frac{\varphi}{3}\right),\quad \varepsilon_2 = -\frac{4\sqrt{7}}{3}\Delta\cos\left(\frac{\pi-\varphi}{3}\right),\quad \varepsilon_3 = -\frac{4\sqrt{7}}{3}\Delta\cos\left(\frac{\pi+\varphi}{3}\right) \tag{4.4.1}$$

and

$$\varphi = \arccos\left(\frac{10D(D^2-9E^2)}{7\sqrt{7}\Delta^3}\right) \tag{4.4.2}$$

The corresponding eigenvectors are given by very formidable formulae and for this reason they will not be quoted here. By calculating them one has to remember that they should block-diagonalize the first order perturbation operator $\hat{V}$. Similarly to the case of $S$=3/2 they are dependent on the sampling field $\boldsymbol{H}$. Likewise, in the limit $\boldsymbol{H} \rightarrow \boldsymbol{0}$ the susceptibility tensor remains dependent on the sampling field orientation. It is given by the following formula

$$\begin{aligned}
\beta^{-1}\chi_{ij}(\boldsymbol{h}) = \frac{\mu_B^2}{Z_2}\Bigg\{ & \frac{B_{ik}(\varepsilon_1,\varepsilon_1)B_{jl}(\varepsilon_1,\varepsilon_1)h_k h_l}{B_{kl}(\varepsilon_1,\varepsilon_1)h_k h_l}\exp(-\beta\varepsilon_1) \\
& + \frac{B_{ik}(\varepsilon_2,\varepsilon_2)B_{jl}(\varepsilon_2,\varepsilon_2)h_k h_l}{B_{kl}(\varepsilon_2,\varepsilon_2)h_k h_l}\exp(-\beta\varepsilon_2) \\
& + \frac{B_{ik}(\varepsilon_3,\varepsilon_3)B_{jl}(\varepsilon_3,\varepsilon_3)h_k h_l}{B_{kl}(\varepsilon_3,\varepsilon_3)h_k h_l}\exp(-\beta\varepsilon_3) \\
& + 2B_{ij}(\varepsilon_1,\varepsilon_2)\exp\left[-\frac{1}{2}\beta(\varepsilon_1+\varepsilon_2)\right]\frac{\sinh\left[\frac{1}{2}\beta(\varepsilon_1-\varepsilon_2)\right]}{\frac{1}{2}\beta(\varepsilon_1-\varepsilon_2)} \\
& + 2B_{ij}(\varepsilon_1,\varepsilon_3)\exp\left[-\frac{1}{2}\beta(\varepsilon_1+\varepsilon_3)\right]\frac{\sinh\left[\frac{1}{2}\beta(\varepsilon_1-\varepsilon_3)\right]}{\frac{1}{2}\beta(\varepsilon_1-\varepsilon_3)} \\
& + 2B_{ij}(\varepsilon_2,\varepsilon_3)\exp\left[-\frac{1}{2}\beta(\varepsilon_2+\varepsilon_3)\right]\frac{\sinh\left[\frac{1}{2}\beta(\varepsilon_2-\varepsilon_3)\right]}{\frac{1}{2}\beta(\varepsilon_2-\varepsilon_3)}\Bigg\}
\end{aligned} \quad (4.4.3)$$

where quantities $B_{ij}(s,t)$ are defined in Section 6.4 and the summation over repeated indices is implied. The susceptibility of the powder sample can be obtained by averaging over all possible orientations of the sampling field as was demonstrated in Section 4.2. Straightforward calculation yields

$$\beta^{-1}\overline{\chi} = \frac{\mu_B^2}{3Z_2}\{B_{ii}(\varepsilon_1,\varepsilon_1)e^{-\beta\varepsilon_1} + B_{ii}(\varepsilon_2,\varepsilon_2)e^{-\beta\varepsilon_2} + B_{ii}(\varepsilon_3,\varepsilon_3)e^{-\beta\varepsilon_3}$$
$$+ 2B_{ii}(\varepsilon_1,\varepsilon_2)\exp\left[-\frac{1}{2}\beta(\varepsilon_1+\varepsilon_2)\right]\frac{\sinh\left[\frac{1}{2}\beta(\varepsilon_1-\varepsilon_2)\right]}{\frac{1}{2}\beta(\varepsilon_1-\varepsilon_2)}$$
$$+ 2B_{ii}(\varepsilon_1,\varepsilon_3)\exp\left[-\frac{1}{2}\beta(\varepsilon_1+\varepsilon_3)\right]\frac{\sinh\left[\frac{1}{2}\beta(\varepsilon_1-\varepsilon_3)\right]}{\frac{1}{2}\beta(\varepsilon_1-\varepsilon_3)} \qquad (4.4.4)$$
$$+ 2B_{ii}(\varepsilon_2,\varepsilon_3)\exp\left[-\frac{1}{2}\beta(\varepsilon_2+\varepsilon_3)\right]\frac{\sinh\left[\frac{1}{2}\beta(\varepsilon_2-\varepsilon_3)\right]}{\frac{1}{2}\beta(\varepsilon_2-\varepsilon_3)}\Bigg\}$$

where the summation over repeated indices is implied. One can readily show that the result is again equivalent to Eq. (4.2.6) where $\chi_{ii}(\boldsymbol{h}_i)$ should be calculated using Eq. (4.4.3).

Figure 6 shows the susceptibility components calculated numerically in the external field $\boldsymbol{H_0} = [1,1,1]$ Oe for $D$=-10 K, $E$=5 K, and $\boldsymbol{g} = \mathrm{diag}(2,2,2)$. The numerical result is consistent with the analytical result given by Eqs. (4.4.3) and (4.4.4) (shown by solid lines). Eq. (4.4.3) works well if both $E \neq 0$ and $D \neq 0$. However, limits $E \to 0$ or $D \to 0$ are rather nontrivial. Therefore, we quote in Section 6.5 the diagonal terms of the susceptibility for these limiting cases.

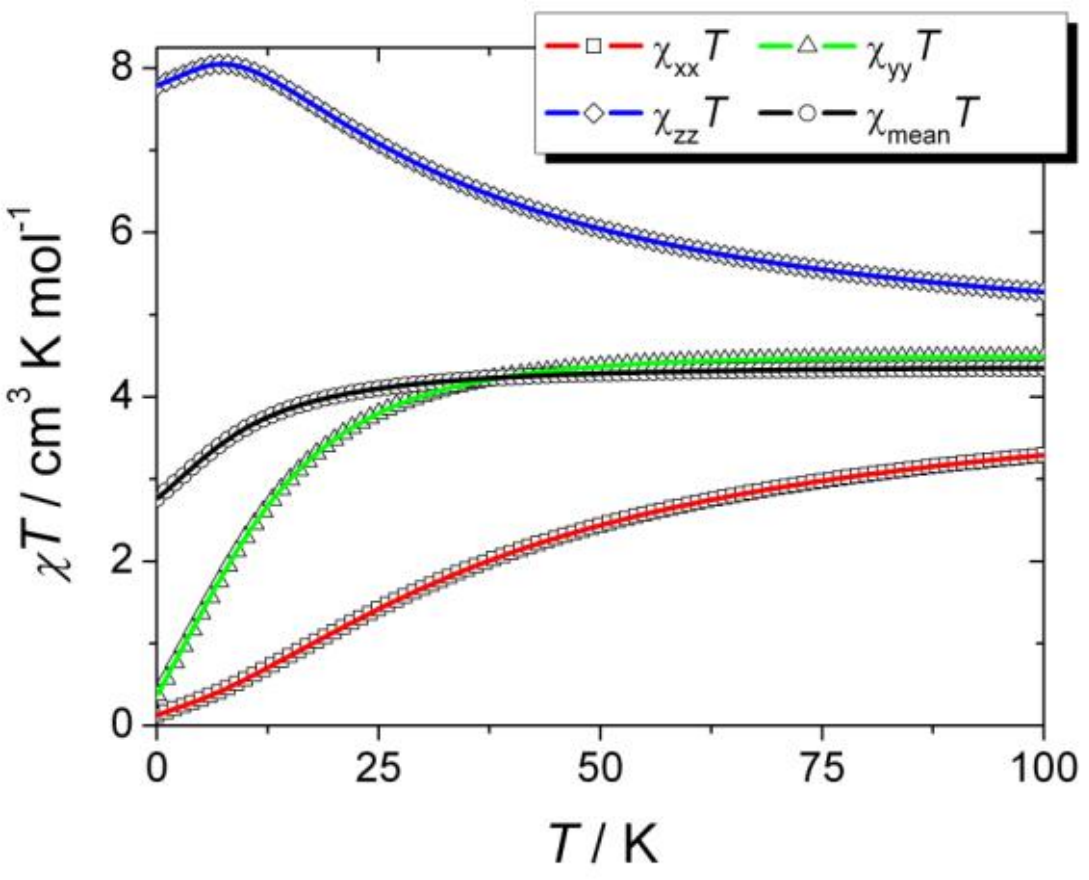

**Fig. 6:** Components of the susceptibility tensor simulated in the model defined by Hamiltonian given in Eq. (4.1) for $S$=5/2, $D$=-10 K, $E$=5 K and $\boldsymbol{g} = \mathrm{diag}(2,2,2)$. The solid lines show the diagonal components of the zero-field susceptibility calculated according to Eq. (4.4.3). The black solid line represents the mean (powder sample) susceptibility calculated according to Eq. (4.4.4). The numerical calculation was performed at the presence of the external magnetic field $\boldsymbol{H_0} = [1,1,1]\,\mathrm{Oe}$.

## 5. Conclusions

The procedure to calculate the magnetic susceptibility and magnetization has been presented and discussed. Let us note that using this procedure in addition to diagonal also the off-diagonal terms of the susceptibility tensor can be obtained, which is of major importance if the experimental single-crystal response should be simulated properly. A simple model of magnetic measurement described by the quantity $m$ was shown to yield only the diagonal components of the susceptibility suggesting the need of experimental set-up allowing to detect the off-diagonal terms. Closed formulae for the full magnetic susceptibility tensor have been obtained for the case of a spin center involving both axial and rhombic zero-field splitting terms. They are ready to be used in the interpretation of magnetic measurements of novel molecular magnets.

## 6. Appendix

**6.1** Eigenvalues and the corresponding eigenvectors of the $\hat{H}_0$ for $S$=1 read

$$E_1^{(0)} = -\frac{2}{3}D \qquad E_2^{(0)} = \frac{1}{3}D - E \qquad E_3^{(0)} = \frac{1}{3}D + E \tag{6.1.1}$$

$$|1\rangle = \begin{pmatrix} 0 \\ 1 \\ 0 \end{pmatrix} \qquad |2\rangle = \frac{1}{\sqrt{2}}\begin{pmatrix} 1 \\ 0 \\ -1 \end{pmatrix} \qquad |3\rangle = \frac{1}{\sqrt{2}}\begin{pmatrix} 1 \\ 0 \\ 1 \end{pmatrix} \tag{6.1.2}$$

**6.2** Doubly degenerate eigenvalues of $\hat{H}_0$ for $S$=3/2 read

$$E_1^{(0)} = \Delta \qquad\qquad E_2^{(0)} = -\Delta \tag{6.2.1}$$

where $\Delta = \sqrt{D^2 + 3E^2}$ , and the corresponding eigenvectors, for which operator $\hat{V}$ is diagonal in the subspaces belonging to the degenerate eigenvalues, are

$$|1,1\rangle = N_1 \begin{pmatrix} (\Delta + D)\left[\lambda_1 + \frac{1}{2}\left(1 + \frac{2D}{\Delta}\right)G_z\right] \\ \frac{\sqrt{3}}{2}E\left[\frac{3E}{\Delta}G_+ + \left(1 - \frac{D}{\Delta}\right)G_-\right] \\ \sqrt{3}E\left[\lambda_1 + \frac{1}{2}\left(1 + \frac{2D}{\Delta}\right)G_z\right] \\ \frac{1}{2}(\Delta + D)\left[\frac{3E}{\Delta}G_+ + \left(1 - \frac{D}{\Delta}\right)G_-\right] \end{pmatrix} \tag{6.2.2}$$

$$|1,2\rangle = N_1 \begin{pmatrix} -\frac{1}{2}(\Delta + D)\left[\frac{3E}{\Delta}G_- + \left(1 - \frac{D}{\Delta}\right)G_+\right] \\ \sqrt{3}E\left[\lambda_1 + \frac{1}{2}\left(1 + \frac{2D}{\Delta}\right)G_z\right] \\ -\frac{\sqrt{3}}{2}E\left[\frac{3E}{\Delta}G_- + \left(1 - \frac{D}{\Delta}\right)G_+\right] \\ (\Delta + D)\left[\lambda_1 + \frac{1}{2}\left(1 + \frac{2D}{\Delta}\right)G_z\right] \end{pmatrix} \tag{6.2.3}$$

$$|2,1\rangle = N_2 \begin{pmatrix} \frac{\sqrt{3}}{2}E\left[\frac{3E}{\Delta}G_- - \left(1 + \frac{D}{\Delta}\right)G_+\right] \\ (\Delta + D)\left[\lambda_2 - \frac{1}{2}\left(1 - \frac{2D}{\Delta}\right)G_z\right] \\ -\frac{1}{2}(\Delta + D)\left[\frac{3E}{\Delta}G_- - \left(1 + \frac{D}{\Delta}\right)G_+\right] \\ -\sqrt{3}E\left[\lambda_2 - \frac{1}{2}\left(1 - \frac{2D}{\Delta}\right)G_z\right] \end{pmatrix} \tag{6.2.4}$$

$$\left|2,2\right\rangle = N_2 \begin{pmatrix} -\sqrt{3}E\left[\lambda_2 - \frac{1}{2}\left(1-\frac{2D}{\Delta}\right)G_z\right] \\ \frac{1}{2}(\Delta+D)\left[\frac{3E}{\Delta}G_+ - \left(1+\frac{D}{\Delta}\right)G_-\right] \\ (\Delta+D)\left[\lambda_2 - \frac{1}{2}\left(1-\frac{2D}{\Delta}\right)G_z\right] \\ -\frac{\sqrt{3}}{2}E\left[\frac{3E}{\Delta}G_+ - \left(1+\frac{D}{\Delta}\right)G_-\right] \end{pmatrix} \quad (6.2.5)$$

where

$$N_1^{-2} = 4\lambda_1\Delta(\Delta+D)\left[\lambda_1 + \frac{1}{2}\left(1+\frac{2D}{\Delta}\right)G_z\right] \quad (6.2.6)$$

$$N_2^{-2} = 4\lambda_2\Delta(\Delta+D)\left[\lambda_2 - \frac{1}{2}\left(1-\frac{2D}{\Delta}\right)G_z\right] \quad (6.2.7)$$

$$\lambda_{1/2} = \frac{1}{2}\sqrt{\left(1\mp\frac{D-3E}{\Delta}\right)^2 G_x^2 + \left(1\mp\frac{D+3E}{\Delta}\right)^2 G_y^2 + \left(1\pm\frac{2D}{\Delta}\right)^2 G_z^2} \quad (6.2.8)$$

$$G_i = \mu_B g_{ij} H_j \qquad G_\pm = G_x \pm iG_y \quad (6.2.9)$$

where the summation over repeated indices is assumed.

**6.3** Eigenvalues and the corresponding eigenvectors of $\hat{H}_0$ for $S$=2 read

$$E_1^{(0)} = 2D \qquad E_2^{(0)} = -D+3E \qquad E_3^{(0)} = -D-3E \quad (6.3.1)$$

$$E_4^{(0)} = 2\Delta \qquad E_5^{(0)} = -2\Delta \quad (6.3.2)$$

$$\left|1\right\rangle = \frac{1}{\sqrt{2}}\begin{pmatrix} 1 \\ 0 \\ 0 \\ 0 \\ -1 \end{pmatrix} \qquad \left|2\right\rangle = \frac{1}{\sqrt{2}}\begin{pmatrix} 0 \\ 1 \\ 0 \\ 1 \\ 0 \end{pmatrix} \qquad \left|3\right\rangle = \begin{pmatrix} 0 \\ 1 \\ 0 \\ -1 \\ 0 \end{pmatrix} \quad (6.3.3)$$

$$|4\rangle = \frac{1}{2\sqrt{\Delta(\Delta+D)}}\begin{pmatrix}\Delta+D\\0\\\sqrt{6}E\\0\\\Delta+D\end{pmatrix} \qquad |5\rangle = \frac{1}{2\sqrt{\Delta(\Delta+D)}}\begin{pmatrix}\sqrt{6}E\\0\\-2(\Delta+D)\\0\\\sqrt{6}E\end{pmatrix} \tag{6.3.4}$$

**6.4** Definition of $B_{ij}(s,t)$ is as follows

$$B_{ij}(s,t) = \frac{1}{b(s)b(t)}[b_1(s,t)g_{xi}g_{xj} + b_2(s,t)g_{yi}g_{yj} + b_3(s,t)g_{zi}g_{zj}] \tag{6.4.1}$$

$$\begin{aligned} b_{1/2}(s,t) = 9\Bigg[&\frac{1}{2}\left(\frac{10}{3}D-s\right)\left(\frac{2}{3}D+s\right)\left(\frac{10}{3}D-t\right)\left(\frac{2}{3}D+t\right) \\ &-5E^2\left(\left(\frac{10}{3}D-s\right)\left(\frac{2}{3}D+t\right)+\left(\frac{10}{3}D-t\right)\left(\frac{2}{3}D+s\right)\right) \\ &\pm 2E\left(\frac{10}{3}D-s\right)\left(\frac{10}{3}D-t\right)\left(\frac{4}{3}D+s+t\right)\Bigg]^2 \end{aligned} \tag{6.4.2}$$

$$\begin{aligned} b_3(s,t) = \Bigg[&\frac{1}{2}\left(\frac{10}{3}D-s\right)\left(\frac{2}{3}D+s\right)\left(\frac{10}{3}D-t\right)\left(\frac{2}{3}D+t\right) \\ &+25E^2\left(\frac{2}{3}D+s\right)\left(\frac{2}{3}D+t\right)-27E^2\left(\frac{10}{3}D-s\right)\left(\frac{10}{3}D-t\right)\Bigg]^2 \end{aligned} \tag{6.4.3}$$

$$b(s) = \left(\frac{10}{3}D-s\right)^2\left(\frac{2}{3}D+s\right)^2 + 18E^2\left(\frac{10}{3}D-s\right)^2 + 10E^2\left(\frac{2}{3}D+s\right)^2 \tag{6.4.4}$$

**6.5** In the limit $E \to 0$ ($S$=5/2) one obtains

$$\begin{aligned} \chi_{xx/yy} &= \frac{\beta\mu_B^2 g_{xx/yy}^2}{Z_3}\left[\frac{9}{2} + 5e^{-4\beta D}\frac{\sinh(2\beta D)}{2\beta D} + 8e^{-\beta D}\frac{\sinh(\beta D)}{\beta D}\right] \\ \chi_{zz} &= \frac{\beta\mu_B^2 g_{zz}^2}{2Z_3}(1+9e^{-2\beta D}+25e^{-6\beta D}) \end{aligned} \tag{6.5.1}$$

where $Z_3 = 2(1+e^{-2\beta D}+e^{-6\beta D})$. In the limit $D \to 0$ the diagonal terms of the susceptibility read

$$\chi_{xx/yy} = \frac{\beta\mu_B^2 g_{xx/yy}^2}{98Z_4}\{225 + 1152\cosh(2\sqrt{7}\beta E) \mp 432\sqrt{7}\sinh(2\sqrt{7}\beta E)$$
$$+ 80\frac{\sinh(\sqrt{7}\beta E)}{\sqrt{7}\beta E}[4\cosh(\sqrt{7}\beta E) \pm \sqrt{7}\sinh(\sqrt{7}\beta E)] + 18\frac{\sinh(2\sqrt{7}\beta E)}{2\sqrt{7}\beta E}\Bigg\} \quad (6.5.2)$$
$$\chi_{zz} = \frac{\beta\mu_B^2 g_{zz}^2}{98Z_4}\left[225 + 18\cosh(2\sqrt{7}\beta E) + 1472\frac{\sinh(2\sqrt{7}\beta E)}{2\sqrt{7}\beta E}\right]$$

where $Z_4 = 2[1 + 2\cosh(2\sqrt{7}\beta E)]$ .

**Acknowledgements**

This work was partially supported by the Polish Ministry of Science and Higher Education within Research Project 0087-B-H03-2008-34.